\documentclass[prb,aps,superscriptaddress,reprint]{revtex4-1}

\usepackage{graphicx}
\usepackage[english]{babel}
\usepackage{SIunits}
\usepackage{color}

\begin{document}

\title{Valley lifetimes of conduction band electrons in monolayer WSe$_2$}

\author{Manfred Ersfeld}
\affiliation{2nd Institute of Physics and JARA-FIT, RWTH Aachen University, 52074 Aachen, Germany}

\author{Frank Volmer}
\affiliation{2nd Institute of Physics and JARA-FIT, RWTH Aachen University, 52074 Aachen, Germany}

\author{Lars Rathmann}
\affiliation{2nd Institute of Physics and JARA-FIT, RWTH Aachen University, 52074 Aachen, Germany}

\author{Luca Kotewitz}
\affiliation{2nd Institute of Physics and JARA-FIT, RWTH Aachen University, 52074 Aachen, Germany}

\author{Maximilian~Heithoff}
\affiliation{2nd Institute of Physics and JARA-FIT, RWTH Aachen University, 52074 Aachen, Germany}

\author{Mark Lohmann}
\affiliation{Department of Physics and Astronomy, University of California, Riverside, California 92521, USA}

\author{Bowen~Yang}
\affiliation{Department of Chemistry and Materials Science $\mathrm{\&}$ Engineering Program, University of California, Riverside, California 92521, USA}

\author{Kenji Watanabe}
\affiliation{Research Center for Functional Materials, National Institute for Materials Science, 1-1 Namiki Tsukuba, Ibaraki 305-0044, Japan}

\author{Takashi~Taniguchi}
\affiliation{International Center for Materials Nanoarchitectonics, National Institute for Materials Science, 1-1 Namiki Tsukuba, Ibaraki 305-0044, Japan}

\author{Ludwig Bartels}
\affiliation{Department of Chemistry and Materials Science $\mathrm{\&}$ Engineering Program, University of California, Riverside, California 92521, USA}

\author{Jing Shi}
\affiliation{Department of Physics and Astronomy, University of California, Riverside, California 92521, USA}

\author{Christoph Stampfer}
\affiliation{2nd Institute of Physics and JARA-FIT, RWTH Aachen University, 52074 Aachen, Germany}
\affiliation{Peter Gr\"unberg Institute (PGI-9), Forschungszentrum J\"ulich, 52425 J\"ulich, Germany}

\author{Bernd Beschoten}
\affiliation{2nd Institute of Physics and JARA-FIT, RWTH Aachen University, 52074 Aachen, Germany}

\begin{abstract}
One of the main tasks in the investigation of 2-dimensional transition metal dichalcogenides is the determination of valley lifetimes. In this work, we combine time-resolved Kerr rotation with electrical transport measurements to explore the gate-dependent valley lifetimes of free conduction band electrons of monolayer WSe$_2$. When tuning the Fermi energy into the conduction band we observe a strong decrease of the respective valley lifetimes which is consistent with both spin-orbit and electron-phonon scattering. We explain the formation of a valley polarization by the scattering of optically excited valley polarized bright trions into dark states by intervalley scattering.
Furthermore, we show that the conventional time-resolved Kerr rotation measurement scheme has to be modified to account for photo-induced gate screening effects. Disregarding this adaptation can lead to erroneous conclusions drawn from gate-dependent optical measurements and can completely mask the true gate-dependent valley dynamics.

\end{abstract}


\maketitle

\section{Formation of a valley polarization}

\begin{figure*}[t]
	\includegraphics[width=\linewidth]{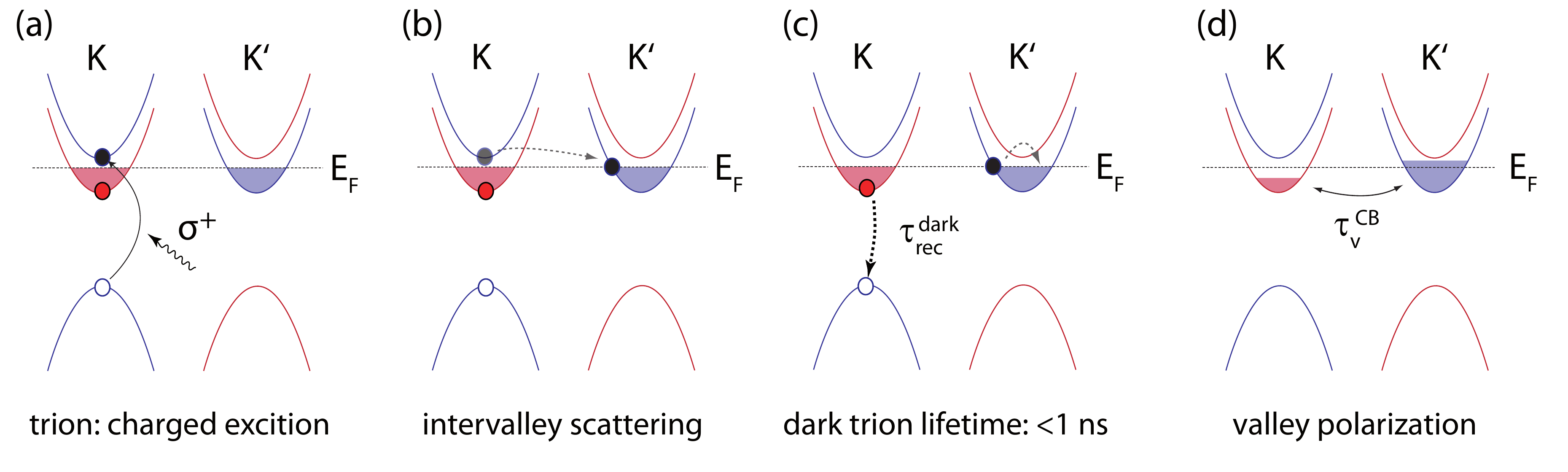}
	\caption{(a) Schematic illustration of a trion state after optical excitation by a circularly polarized laser pulse in monolayer WSe$_2$ where the Fermi energy $E_F$ is tuned into the conduction band. (b) Intervalley scattering of the photoexcited electron into the lower conduction band of the K'-valley forming a dark trion state. (c) Spin-flip recombination of the dark trion which reduces the number of electrons in the K-valley while the number of electrons in the K'-valley gets increases by the same amount. (d) Net valley polarization which is formed after the recombination of all electron-hole pairs. The conduction band valley lifetime
	$\tau_{V}^{CB}$ can be measured by time-resolved Kerr rotation.}
	\label{Fig1}
\end{figure*}

Monolayers of transition metal dichalcogenides have direct band gaps and exhibit spin-polarized valleys which allow to create valley polarized excitons by circularly polarized light.\cite{NatPhys.10.343,RevModPhys.90.021001} This excitation process relies on valley-dependent optical selection rules, which is illustrated in Fig.~\ref{Fig1}a for monolayer WSe$_2$. Here, a $\sigma^+$ photon creates an electron-hole pair at the K-valley by promoting an electron from the valence band into the upper conduction band. As the Fermi level is tuned into the conduction band, the electron-hole pair can easily bind an extra electron with opposite electron spin from the lower conduction band to form a trion.\cite{NatureComm.5.3876,NatSciRev.2.57} While the formation of the trions results in an imbalance of the electron occupation of the conduction band between the K- and K' valley, it will directly be lost if the trion recombines by a direct optical transition. A net valley polarization of free conduction band carriers after recombination of the photoexcited electron and holes states can, however, be created if for example one of the electrons scatter into the other valley. As illustrated in Fig.~\ref{Fig1}b, the photo-excited electron can easily scatter from the upper conduction band at the K-valley into the lower conduction band at the K'-valley which exhibit the same spin orientation. This process creates a dark trion state. \cite{RevModPhys.90.021001,PhysicalReviewMaterials.2.014002} The direct recombination of the electron in the K'-valley with the hole in the K-valley is momentum-forbidden and requires an interaction with a phonon. We note that this recombination does not yield a valley polarization as the other electron at the K-valley goes back into the Fermi see as a free electron after recombination. In contrast, a net valley polarization can be created if the latter electron recombines with the a hole in the same K-valley (see Fig.~\ref{Fig1}c).\cite{PhysRevLett.119.047401} The previously photo-excited electron will therefore become a free charge carrier in the K'-valley. The spin-flip recombination in the K-valley reduces the number of electrons in the K-valley while the number of electrons in the K'-valley gets increased by the same amount. We note that the number of spin-flip events is limited by the initial number of free electrons in the K-valley. When tuning the Fermi energy into the conduction band, the valley polarization is thus expected to increase with the number of free charge carriers. As the Kerr rotation angle is a direct measure of the valley polarization, it is straightforward to demonstrate that a valley polarization of free carriers is indeed created by the above scattering process.

\begin{figure*}[th]
	\includegraphics[width=\linewidth]{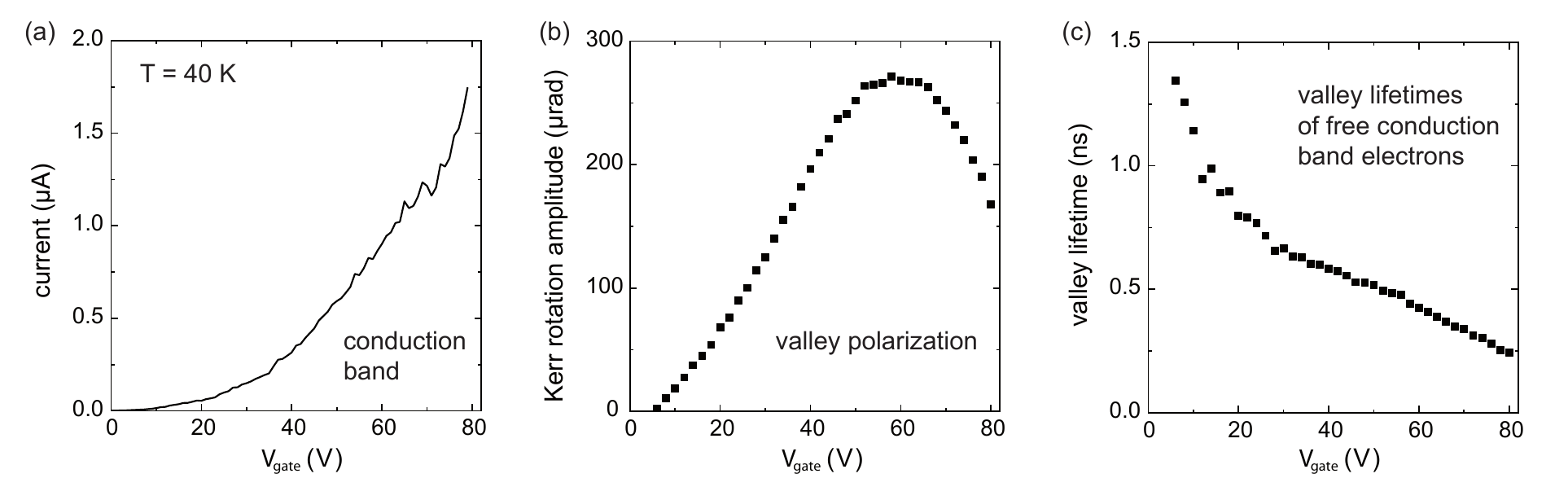}
	\caption{(a) Gate dependent electrical transport measurement taken on a monolayer of WSe$_2$ at 40~K. Above the conduction band edge, the current increases with gate voltage, i.e. charge carrier density. (b) Gate dependent Kerr rotation amplitude which is a measure of the valley polarization of free conduction band electrons. (c) Gate dependent valley lifetime of free conduction band electrons.}
	\label{Fig2}
\end{figure*}

\section{Valley lifetimes of free conduction band electrons}

In order to measure the valley lifetime $\tau_{V}^{CB}$ of free conduction band electrons (see Fig.~\ref{Fig1}d), we combine time-resolved Kerr rotation (TRKR) and electrical transport measurements on monolayer WSe$_2$ which is encapsulated by hexagonal boron nitride (hBN) and electrically contacted by graphite electrodes. The device can be tuned into the conduction band by electrostatic gating as seen by the gate-voltage dependent increase of the current (Fig.~\ref{Fig2}a). Time-resolved Kerr rotation traces has been recorded for photon energies resonantly exciting trion states.\cite{PhysRevB.95.235408,Opt.Mater.Express.10.1273} In Figs.~\ref{Fig2}b,c, we show the gate-dependent Kerr rotation amplitudes and the respective lifetimes at 40~K. When entering the conduction band at around +10~V gate voltage there is a linear increase in the Kerr rotation amplitude. This behavior is consistent with the above decay process of dark trions, which results in a net valley polarization after a full recombination of all optically excited excitons. The saturation and decrease of the Kerr rotation amplitude towards very high gate voltages can be explained by the filling of both the upper, spin-inverted conduction bands at the K-valleys and the spin-degenerated bands at the Q-valleys ($\Lambda$-valleys), which will reduce the overall net valley polarization.\cite{RevModPhys.90.021001} The corresponding valley lifetimes are largest near the bottom of the conduction band and strongly decreases with increasing gate voltage, i.e. charge carrier densities. Such a decrease is expected from both spin-orbit and electron-phonon scattering mechanisms.\cite{NanoLetters.17.4549,PhysRevB.87.245421}

\begin{figure*}[th]
	\includegraphics[width=\linewidth]{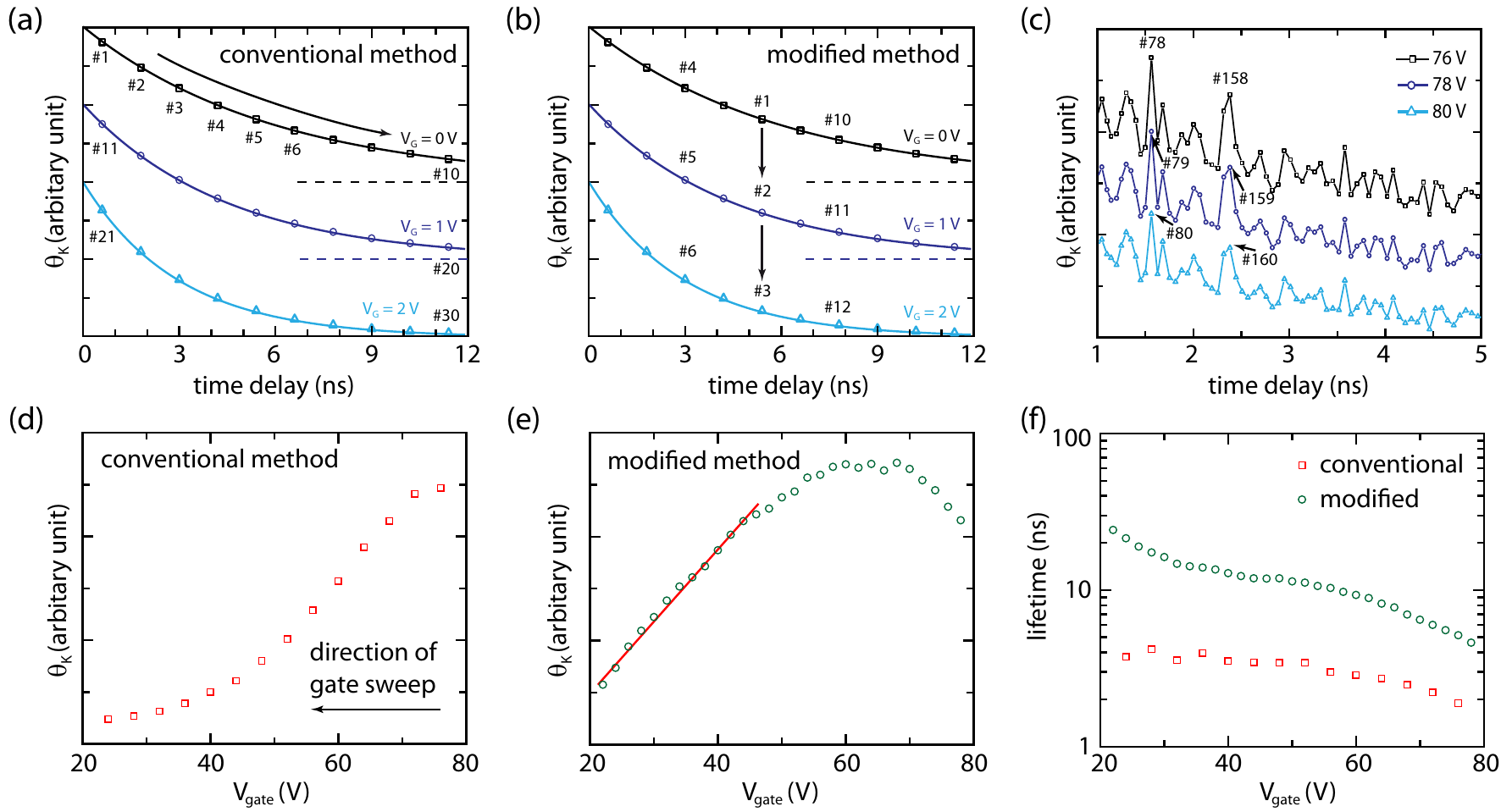}
	\caption{(a) Schematic of the conventional TRKR measurement technique in which the time delay between the pump and probe laser pulse is step by step increased from $\Delta t =$ 0~ns to 12~ns before the next TRKR trace is measured for the next gate voltage (see numbering of data points). As the photo-induced screening of the gate-electric field varies on laboratory time scales, subsequent data point which are nominally measured at the same gate-voltage are in fact recorded at different charge carrier densities. (b) Schematic of the modified TRKR measuring technique in which a full gate-sweep is done for each time delay (see numbering of data points). (c) Because the time delays in this modified technique are distributed randomly, a change in device properties over laboratory time manifests itself in an apparent "noise" signal seen when comparing different gate voltages. (d) to (f) Comparison between the conventional and the modified measurement technique. The two gate-dependent TRKR measurements were done right after each other but yield significantly different amplitudes, lifetimes and overall gate-dependent trends.}
	\label{Fig3}
\end{figure*}

\section{Modification of time-resolved Kerr rotation measurement technique to account for photo-induced gate screening effects}

In this section we discuss that the time-resolved Kerr rotation measurement scheme has to be modified as soon as optical measurements such as photoluminescence (PL) or magneto-optical probing show hysteresis effects during electrostatic gating. The hysteresis can result from a screening effect of the gate electric field by photo-excited charged defects in the dielectric layer.\cite{NatureNanotechnology.9.348,ACSAppl.Mater.Interfaces.8.9377,Nanoscale.11.7358,Volmer2020Jul} We demonstrate that disregarding the adaptation of the measurement scheme can lead to erroneous conclusions drawn from gate-dependent measurements and can completely mask the true gate-dependent valley dynamics.

First, we discuss the "conventional" measurement technique in which the gate-voltage is set to a fixed value and then the full TRKR curve is measured by varying the time delay $\Delta t$ between pump and probe pulses over laboratory time. This is depicted in the schematic of Fig.~\ref{Fig3}a. For the sake of simplicity we assume that for each TRKR curve the Kerr rotation amplitude $\Theta_K$ is measured at ten successive time delays before repeating the measurement for the next gate voltage. Here, data point \#1 is measured right after setting the gate voltage, whereas the data point \#10 may be recorded minutes later (this is the measurement time of a real TRKR curve consisting of up to several hundreds of data points).

With this measurement technique, once the gate-voltage is set to a certain value, the photo-induced screening of the gate-electric field will start to a change the gate-induced charge carrier densities over time.\cite{Volmer2020Jul} Although all data points within a Kerr rotation curve are therefore measured at a nominal identical gate voltage, each single point is effectively measured at a different Fermi level position.

To avoid this problem, our adapted measurement technique swaps the sequence in which the two parameters of gate voltage and delay time are set: First, a fixed time delay is set and then the Kerr rotation signal is measured for a full gate voltage sweep (see numbering of data points in Fig.~\ref{Fig3}b). To achieve the maximum comparability between different measurements, the same gate sweep direction and velocity should be used for all TRKR, PL, and transport measurements.\cite{Volmer2020Jul} This ensures that the photo-doping effect impacts all measurements in the same way.

With this modified measurement technique, each data point of a TRKR curve has the same "history", i.e., the temporal course of both setting and staying at previous gate voltages is exactly the same and each data point is measured exactly after the same amount of time after setting the corresponding gate voltage. Assuming that the impact of the photo-induced screening is reproducible, all data points of a specific TRKR curve are therefore measured at the same charge carrier density.

A further important aspect of the modified Kerr measurement technique is a randomized sequence of the measured time delays (see numbering of data points in Fig.~\ref{Fig3}b). This randomization is important to identify changes in device properties on laboratory time-scales, which are especially present within the first few hours after the cool-down of a device. These temporal changes are most likely due to a degassing of adsorbed molecules on top of the device under laser illumination.\cite{APL.114.172106}

Within the "conventional" measurement scheme, such a temporal change in device properties over laboratory time may be seen as an exponentially decaying signal in a TRKR measurement and therefore can lead to a misinterpretation of the measurement. Instead, in the modified measurement technique the change over laboratory time is randomly distributed over all measured delay times. Therefore, such a change in device properties can easily be identified as an apparent "noise" signal in delay scans at different gate voltages.

This is shown in Fig.~\ref{Fig3}c which depicts TRKR curves at different gate voltages measured right after cooling down the device. For better visibility, the TRKR curves are vertically shifted to each other. The apparent noise in these curves is almost identical, showing that in fact the device slowly changed over time. The change in Kerr rotation amplitude due to the photo-induced effects is therefore much more pronounced than the actual noise.

It should be noted that within the randomized sequence of time delays, the first two measurements were done at around 1.6~ns and 2.4~ns. Therefore, at these two time delays peaks occur in the TRKR curves as the time-dependent change in device properties is most pronounced at the start of a measurement. It is good practice to discard such measurements and to wait until the device response has settled to such an extent that there is a truly random noise between different traces.

In Figs.~\ref{Fig3}d to \ref{Fig3}f the two measurement techniques are compared. The two depicted sets of gate-dependent TRKR measurements were performed right after each other but yield significantly different amplitudes, lifetimes and overall gate-dependent trends. In case of the conventional method, the first TRKR curve was measured at a gate voltage close to 80~V and then the voltage was continuously decreased to 20~V. Therefore, the Kerr rotation amplitude almost decays exponentially in the direction of the gate sweep (see Fig.~\ref{Fig3}d). This clearly indicates the exponentially decaying change in device properties caused by the photo-induced effects. In contrast, with the modified measurement technique a linear increase in polarization with increasing gate voltage can be observed (see green line in Fig.~\ref{Fig3}e) which we could recently assign to a valley polarization of free charge carriers.\cite{NanoLetters.20.3147}

Not only the gate-dependent Kerr rotation amplitudes but also the extracted lifetimes are significantly different between both measurement techniques as depicted in Fig.~\ref{Fig3}f. The apparent lifetimes in case of the conventional technique are much shorter than the ones obtained by the modified technique. This is due to the fact that for the conventional technique the exponentially decreasing change in device properties over laboratory time is projected onto the sweep velocity of the delay-time: A slow measurement of a time-delay trace would project the largest change in device properties into the first few measured time delays, resulting in an apparent Kerr signal with short lifetime. On the other hand, if the time-delay trace is measured quickly the same change in device properties would be distributed over a larger span of measured time-delays, therefore yielding an apparent Kerr signal with a longer lifetime.

\

{\bf Acknowledgements:}
 This project has received funding from the European Union's Horizon 2020 research and innovation programme under grant agreement No. 881603 (Graphene Flagship), the Deutsche Forschungsgemeinschaft (DFG, German Research Foundation)  under  Germany’s  Excellence  Strategy - Cluster of Excellence Matter and Light for Quantum Computing (ML4Q) EXC 2004/1 - 390534769 and by the Helmholtz Nanoelectronic Facility (HNF) at the Forschungszentrum J\"ulich \cite{HNF}. K.W. and T.T. acknowledge support from the Elemental Strategy Initiative conducted by the MEXT, Japan, Grant Number  JPMXP0112101001, JSPS KAKENHI Grant Number JP20H00354 and the CREST(JPMJCR15F3), JST.

\end{document}